\documentstyle[12pt,aps]{revtex} 

\begin{document}

\title
{DEPHASING OF LOCAL VIBRATIONS 
IN A PLANAR LATTICE  OF ADSORBED MOLECULES}

\author
{
V. M. Rozenbaum{$^{1}$} and I. V. Kuzmenko{$^{2}$}
\footnote{E-mail: {$^{1}$}vrozen@mail.kar.net, 
                  {$^{2}$}tanya@ap3.bitp.kiev.ua}
}

\address
{Institute of Surface Chemistry\\ 
National Academy of Sciences of Ukraine\\ 
Prosp. Nauki 31, Kyiv-22, 252022 Ukraine}

\maketitle

\vspace{5mm}

\begin{abstract}
        We investigate anharmonically coupled high-- and 
low--frequency excitations in a planar lattice of adsorbed 
molecules interacting with phonons of a crystal. Dephasing 
of high--frequency local vibrations by low--frequency resonance 
modes is described in terms of temperature Green's function. 
The equations obtained are solved, first, with a small ratio 
of the anharmonic coupling coefficient for high-- and 
low--frequency modes to the resonance width, and second, 
in the low--temperature limit. High--frequency spectral line 
positions and widths depend on dispersion laws and resonance 
mode lifetimes. It is shown that lateral interactions of 
low--frequency modes of adsorbed molecules can lead to 
a significant narrowing of high--frequency spectral lines, 
which is consistent with experimental data.
\end{abstract}

\vspace{10mm}

        A considerable progress in the understanding of 
the broadening mechanism for a high--frequency local vibration 
spectral line of an impurity molecule has been achieved due to 
studies on the self-scattering (dephasing) of this vibration 
arising from a biquadratic anharmonic coupling between high-- 
and low--frequency molecular modes~[1--3]. At a sufficiently 
high frequency of the local vibration, other anharmonic 
coupling types do not contribute to the spectral line 
broadening~[3--5] with the exception of additionally 
renormalizing the biquadratic anharmonicity coefficient~[6]. 
In this context, of significance is that the low--frequency 
mode acquires a resonance nature due to its interaction with 
the quasicontinuous spectrum of crystal lattice phonons.  
In the treatment of the self-scattering effect, the dephasing 
model~[4,7,8] proved fruitful, as it afforded a number of exact 
solutions~[9--12] and hence a description of spectral line shapes 
for local vibrations. 

        If the amount of adsorbed molecules suffices to form 
a monolayer, the dephasing model should take into account 
collectivization of vibrational molecular modes~[13--16]. 
As a rule, high-- and low--frequency lattice excitations 
are respectively related to stretch and deformation 
(translational or orientational) molecular vibrations.

        In this paper, a dephasing model for local vibrations 
is investigated for a system of anharmonically coupled high-- 
and low--frequency molecular vibrations, with crystal phonons 
included. If translational symmetry is characteristic of 
the system considered, the low--frequency contribution to 
its Hamiltonian can be introduced in a diagonal form with 
respect to the two-dimensional wave vector {${\bf K}$} 
belonging to the first Brillouin zone of a planar lattice~[17]. 
Then the total Hamiltonian is written as a sum of harmonic and 
anharmonic contributions:

\begin{equation}
H_{tot}=H+H_{A}, \ \ \ H=\sum_{\bf K} H_{\bf K},  
\end{equation}

\begin{equation}
H_{\bf K}=\hbar\Omega_{\bf K}a^{+}_{\bf K}a_{\bf K}+
\hbar\omega_{\bf K}b^{+}_{\bf K}b_{\bf K}+
\hbar\sum_{\nu}\omega_{{\bf K},\nu} \beta^{+}_{{\bf K},\nu}
\beta_{{\bf K},\nu}+
\hbar\sum_{\nu}(\chi_{{\bf K},\nu}b^{+}_{\bf K}\beta_{{\bf K},\nu}
     +H.c.),
\end{equation}

\begin{equation}
H_{A}=\frac {\hbar\gamma}{N}
\sum_{\bf KK'K''}a^{+}_{\bf K}a_{\bf K'}b^{+}_{\bf K''}
b_{\bf K-K'+K''},
\end{equation}
where the terms in the expression (2) respectively represent 
high--  and low--frequency vibrations of a molecular lattice, 
crystal phonons, and the harmonic interactions between phonons 
and low--frequency modes. In the above relation, quantum states 
of phonons of the admolecular lattice are characterized by 
the surface--parallel wave vector ${\bf K}$, whereas the quantum 
numbers of substrate phonons are indicated by the couple 
of indices ${\bf K}$ and $\nu$.  The latter accounts for 
the polarization of a quasi--particle and its motion in 
the surface--normal direction; it implicitly reflects both 
the atomic arrangement in the crystal unit cell and the position 
of the admolecular lattice, as a whole, relative to the crystal 
surface. The utility of introduction of the complex index 
{$\nu$} was substantiated previously~[17].

        As a rule, the density of states for molecular lattice 
vibrations is negligible as compared to that for crystal phonons. 
Therefore, the ${\bf K}$--mode of a molecular lattice interacts 
with the crystal phonons specified by the same wave vector 
${\bf K}$. Besides, the low--frequency collective mode 
{${\omega}_{\bf K}$} of adsorbed molecules can be considered 
as a resonance vibration with the renormalized frequency 
${\tilde\omega}_{\bf K}$ and the inverse lifetime 
${\eta}_{\bf K}\left ( {\tilde\omega}_{\bf K} \right )$
[12,17].

        The response of the system concerned to an external 
electromagnetic field is conveniently described in terms of 
double-time Green's function (GF) which can be introduced in 
a variety of representations~[18--22]. In what follows we will 
involve the representation in Matsubara's frequency space~[18] 
which is accepted in the theory of anharmonic crystals~[23] and 
provides a number of exact solutions in the case of a single 
adsorbed molecule~[10,11]. In this approach, the spectral line 
shape for high--frequency vibrations can be determined as follows:

\begin{equation}
L(\omega)=\frac{\hbar\beta}{\pi}
{\mbox{Im}} G(a_{\bf 0}a^{+}_{\bf 0},\omega),
\ \ \ {\beta}=(k_{B}T)^{-1},
\end{equation}

\begin{equation}
G(a_{\bf 0}a^{+}_{\bf 0},\omega)=\frac{1}{\beta}
\int \limits_{-\infty}^{+\infty}
d{\tau}G(a_{\bf 0}a^{+}_{\bf 0},{\tau})e^{i\omega \tau}, 
\end{equation}

\begin{equation}
G(a_{\bf 0}a^{+}_{\bf 0},\tau)=
\frac{ \Bigl < \hat\tau \tilde a_{\bf 0}(\tau) 
\tilde a^{+}_{\bf 0}(0) \hat S(\beta) \Bigr >_{0} }
{\Bigl < \hat S(\beta) \Bigr >_{0}},
\end{equation}
where

\begin{eqnarray}
\hat S(\beta) & = & \hat S(\beta,0),  \nonumber \\
\hat S(\tau,{\tau}_{0}) & = & \hat\tau \exp \left [ 
-\int \limits_{{\tau}_{0}}^{\tau} d{\tau}'{\tilde H}_{A} ({\tau}')
\right ], 
\end{eqnarray}

\begin{equation}
\tilde A = e^{\hat H \tau} \hat A e^{- \hat H \tau}, \ \ \
\Bigl < ... \Bigr >_{0} =\frac{Sp\left\{ e^{-\beta H}...\right\} }
{Sp\left\{e^{-\beta H}\right\} },
\end{equation}
and $\hat A$ is an arbitrary operator function.

Expanding the right side of Eq.~(6) in powers of the anharmonic 
coupling coefficient $\gamma$, the temperature GF (6) is expressed 
in terms of the GF's of high-- and low--frequency vibrations of 
a planar lattice in the harmonic approximation:

\begin{eqnarray}
G^{(0)}_{\bf K}(\tau) & = & 
\Bigl <\hat \tau a_{\bf K}(\tau) a^{+}_{\bf K}(0) \Bigr >_{0} 
\nonumber \\
 & = & \Bigl\{ [n({\Omega}_{\bf K})+1]\theta(\tau)+
                n({\Omega}_{\bf K})\theta(-\tau)\Bigr\}
       \exp (-\hbar{\Omega}_{\bf K}\tau),
\end{eqnarray}

\begin{eqnarray}
g^{(0)}_{\bf K}(\tau) & = & 
\Bigl <\hat \tau b_{\bf K}(\tau)b^{+}_{\bf K}(0)\Bigr >_{0} 
\nonumber \\
 & = & \int\limits_{-\infty}^{\infty}d\omega{\Re}_{\bf K} (\omega)
\Bigl\{ [n(\omega)+1]\theta(\tau)+n(\omega)\theta(-\tau)\Bigr\}
\exp (-\hbar\omega\tau),
\end{eqnarray}
where

\begin{equation}
n(\omega)=\left [e^{\beta \hbar \omega}-1 \right]^{-1}
\end{equation}
is the is Bose--factor and 

\begin{equation}
{\Re}_{\bf K}(\omega)=
\frac{\tilde\eta_{\bf K} (\omega)}
{\left ( \omega-\omega_{\bf K}- \tilde P_{\bf K}(\omega)\right )^{2}
+\pi^{2}{\tilde \eta}^{2}_{\bf K}(\omega)} 
\end{equation}
is the resonance spectral function of low--frequency 
vibrations~[12,24],

\begin{equation}
\tilde \eta_{\bf K}(\omega)=\sum_{\nu}\vert \chi_{K,\nu}\vert^{2}
\delta(\omega-\omega_{{\bf K},\nu}), \ \ \ \
\tilde P_{\bf K}(\omega)=\int \limits_{-\infty}^{\infty}
\frac{\tilde \eta_{\bf K}(\bar \omega)d\bar \omega}
     {\omega-\bar \omega}.
\end{equation}

The real and imaginary parts of the pole in expression (12) 
define the renormalized  frequency $\tilde \omega_{\bf K}$  
and the inverse lifetime $\eta_{\bf K}=2\pi\tilde\eta_{\bf K}
(\tilde\omega_{\bf K})$ of the resonance vibration.

        Let us take advantage of the inequality 
$\beta\hbar\Omega_{\bf K}>>1$ permitting neglect of 
the terms of the order $\exp (-\beta\hbar\Omega_{\bf K})$. 
Then the trace taken over high-- and low--frequency modes,  
Sp\{...\}, is reduced to that for low--frequency modes, with 
all high--frequency vibrations considered only for the ground 
state. The resulting temperature GF (6) takes the form:

\begin{equation}
G(a_{\bf 0}a^{+}_{\bf 0},\tau)=
\Bigl < \tilde a_{\bf 0}(\tau)\hat S(\tau) 
\tilde a^{+}_{\bf 0}(0) \Bigr >_{0}, \ \ \
\beta > \tau > 0.
\end{equation}
        
Thus, provided the inequality $\beta\hbar\Omega_{\bf K}>>1$ 
is valid, the GF defined above can be written in the form 
involving no high--frequency mode operators:

\begin{equation}
G(a_{\bf 0}a^{+}_{\bf 0},\tau)=
\exp\left(\hbar\Omega_{\bf 0}\tau\right)
\Bigl < \hat s_{\bf 00}(\tau)\Bigr >_{0},
\end{equation}

\begin{equation}
\hat s_{\bf KK'}(\tau)=\sum_{n=0}^{\infty}(-\hbar\gamma)^{n}
\int\limits_{0}^{\tau}d\tau_{1}...
\int\limits_{0}^{\tau_{n-1}}d\tau_{n}
\sum_{{\bf K}_{1}...{\bf K}_{n-1}}
\tilde B_{{\bf K}      {\bf K}_{1}} (\tau_{1})
\tilde B_{{\bf K}_{1}  {\bf K}_{2}} (\tau_{2})...
\tilde B_{{\bf K}_{n-1}{\bf K}'   } (\tau_{n}),
\end{equation}

\begin{equation}
\tilde B_{\bf K}(\tau)=\frac{1}{N}
\exp\left [ \hbar (\Omega_{\bf K}-\Omega_{\bf K'})\right]
\sum_{\bf K''} b^{+}_{\bf K''- K} (\tau) b_{\bf K''- K'} (\tau).  
\end{equation}

Representing average value (15) as

\begin{equation} 
\Bigl <\hat s_{\bf 00}(\tau)\Bigr >_{0}=e^{F(\tau)} 
\end{equation}
and differentiating the left and the right sides of 
expression~(18) with respect to $\tau$, we arrive at: 

\begin{equation}
\frac {dF(t)}{dt}=-\hbar\gamma 
\sum_{\bf K}
\frac{\Bigl < \tilde B_{\bf 0K} (\tau)
\hat s_{\bf K0}(\tau)\Bigr >_{0} } 
{\Bigl <\hat S_{\bf 00}(\tau)\Bigr >_{0}}.
\end{equation}

Eq.~(19) allows the line shape to be determined in 
the approximation of the small anharmonic coefficient 
$\gamma$, as well as in the low--temperature limit.

        Importantly, the three-time GF

\begin{equation}
g_{\bf KK'}(\tau,{\tau}',{\tau}'')=
\frac{\Bigl <\hat \tau \tilde b_{\bf K}(\tau) 
\tilde b^{+}_{{\bf K}'}({\tau}')
\hat s_{{\bf K}-{\bf K}',0}({\tau}'')\Bigr >_{0}} 
{\Bigl <\hat s_{\bf 00}(\tau)\Bigr >_{0}} 
\end{equation} 
satisfies the Dyson--like equation with the assumption 
of non--interacting high--frequency collective vibrations 
($\Omega_{\bf K}=\Omega_{\bf 0}$):

\begin{equation} 
g_{\bf KK'}(\tau,{\tau}',{\tau}'')= 
g^{(0)}_{\bf K}(\tau-\tau^{\prime})\delta_{\bf KK'}+
\frac{\hbar\gamma}{N}\sum_{{\bf K}_{1}}
\int\limits_{0}^{{\tau}''} d\tau_{1}g^{(0)}_{\bf K}(\tau-\tau_{1})
g_{{\bf K}_{1},{\bf K}'}(\tau_{1},\tau^{\prime},{\tau}'').
\end{equation}
(the resonance GF $g^{(0)}_{\bf K}(\tau)$ in the harmonic 
approximation is determined by the expression~(10)). 
Equation~(21) takes into account intermolecular interactions 
of low--frequency resonance modes and thus generalizes 
the corresponding equations derived in papers~[10,11].

        To invoke the perturbation theory for small anharmonic 
coupling coefficient, we use the Wick theorem for the coupling 
of the creation and annihilation operators of low--frequency 
modes in expression (19). Retaining the terms of the orders 
$\gamma$ and ${\gamma}^{2}$ we are led to the following 
expressions for the shift $\Delta\Omega$ and the width 
$2\Gamma$ of the high--frequency vibration spectral line:

\begin{equation}
\Delta\Omega=\frac{\gamma}{N}
\sum_{\bf K}\int\limits_{-\infty}^{\infty}d\omega
   n(\omega) {\Re}_{\bf K} (\omega), \ \ \ \
\end{equation}

\begin{equation}
2\Gamma=\frac{2\pi\gamma^{2}}{N^2}
\sum_{\bf KK'}\int\limits_{-\infty}^{\infty}d\omega
  n(\omega) {\Re}_{\bf K+K'} (\omega) 
  {\Re}_{\bf K'} (\omega+\Omega_{\bf 0}-\Omega_{\bf K}),
\end{equation}
where $n \left ( \omega \right ) $ and ${\Re}_{\bf K} (\omega)$ 
are respectively determined by relations (11) and (12). It should 
be noted that expressions (22) and (23) describe a single 
Lorentz--like spectral line of local vibrations, with its position 
and width dictated by dispersion laws and lifetimes for resonance 
low--frequency modes.

        The analytical description of high--frequency line shapes 
becomes possible in the low--temperature limit, i.e., at 
$n(\omega_{\bf K})\approx\exp(-\beta\hbar\omega_{\bf K})<<1$, 
which represents an experimentally important case. In this 
situation, the Wick coupling for the operators of low--frequency 
modes in expression (19) involves only the terms in which 
the annihilation operator is to the left of the creation 
operator in all but one operator pair. Then Eq. (19) can 
be written as:
 
\begin{eqnarray}
\frac{dF(\tau)}{d\tau} & = & -\frac{\hbar\gamma}{N}\sum_{\bf K}
d\omega \Re_{\bf K} ( \omega ) n( \omega )
\sum_{n=0}^{\infty} \left ( -\hbar\gamma \right )^{n} 
\int\limits_{0}^{\tau}d{\tau}_{1} 
    {\varphi}_{\bf K} \left (\omega, {\tau}-{\tau}_{1}\right )
\nonumber\\
 & & \times
\int\limits_{0}^{{\tau}_{1}}d{\tau}_{2} 
    {\varphi}_{\bf K} \left (\omega, {\tau}_{2}-{\tau}_{1}\right )
\ldots
\int\limits_{0}^{{\tau}_{n-1}}d{\tau}_{n} 
    {\varphi}_{\bf K} \left (\omega, {\tau}_{n}-{\tau}_{1}\right )
\end{eqnarray}
where

\begin{eqnarray}
{\varphi}_{\bf K} \left (\omega, {\tau}\right ) & = & 
\frac{1}{N}\sum_{\bf K'}\int\limits_{-\infty}^{\infty}
d{\omega}'
\exp\left [ \hbar \left ( 
    {\Omega}_{0}-{\Omega}_{{\bf K}-{\bf K'}}+
    {\omega}-{\omega}' \right ) \tau \right ]
\nonumber\\
 & \approx & \frac{1}{N}\sum_{\bf K'}
\exp\left [ \hbar \left ( 
    {\Omega}_{0}-{\Omega}_{{\bf K}-{\bf K'}}+
    {\omega}-{\omega}_{\bf K'}+
    i\pi{\tilde\eta} ({\omega}_{\bf K'})
    \right ) \tau \right ]
\end{eqnarray}
On integrating Eq.~(24) over the variables ${\tau}_{1}$, 
${\tau}_{2}$, $\ldots$~, ${\tau}_{n}$, we obtain:

\begin{eqnarray}
\frac{dF( \tau )}{d\tau} & = & -\frac{\hbar\gamma}{2\pi N}
\sum_{\bf K}\int\limits_{-\infty}^{\infty}d\omega
{\Re}_{\bf K}( \omega ) n( \omega )
\int\limits_{-\infty}^{\infty} dy 
\frac{e^{\tau (\lambda+iy)}}{\lambda+iy}
\nonumber\\
 & & \times\left [ 1+\hbar\gamma\int\limits_{-\infty}^{\infty}
{\varphi}_{\bf K} \left (\omega, x \right )
e^{-x(\lambda+iy)} \right ]^{-1}.
\end{eqnarray}
Then expression~(26) in the Markov approximation 
$\left [\hbar{\tilde\eta}_{\bf K}( \omega )\tau, \ 
\hbar\gamma\tau >> 1 \right ]$ can be reduced as follows: 

\begin{equation}
\frac{dF(\tau )}{d\tau}=-\frac{\hbar\gamma}{N}
\sum_{\bf K}\frac{n \left ({\omega}_{\bf K} \right )}
{1-\left (\gamma/N \right )\sum_{\bf K'}
\left \{ {\alpha}_{\bf KK'}+
i\pi\left [{\tilde\eta}_{\bf K}({\omega}_{\bf K}+
{\tilde\eta}_{\bf K'}({\omega}_{\bf K'}  \right ] \right \}^{-1}},
\end{equation}
where

\begin{equation}
{\alpha}_{\bf KK'}={\Omega}_{0}-{\Omega}_{{\bf K}-{\bf K'}}+
    {\omega}_{\bf K}-{\omega}_{\bf K'}.
\end{equation}
As seen, the spectral line of high--frequency local vibrations 
is of the Lorentz--like shape: 

\begin{equation}
L\left ( \omega \right ) = -\frac{1}{\pi}{\mbox{Im}}
\frac{1}{\omega-{\Omega}_{0}-W}, \ \ \ 
W = -\frac{dF( \tau )}{d\tau},
\end{equation}
with the shift $\Delta\Omega={\mbox{Re}}W$ and the width 
$2\Gamma=-2{\mbox{Im}}W$.

        To estimate the lateral interaction effect 
on basic spectral characteristics of high--frequency 
local vibrations, assume that the density of states for 
the low--frequency band is described by the step function:

\begin{equation}
{\rho}_{\varphi}\left ( \omega \right ) =
\Biggl\lbrace { \begin{array}{c}
{\Delta\omega}^{-1}, \ \ \ 
     {\omega}_{0}<{\omega}<{\omega}_{0}+{\Delta\omega}, \\
0, \ \ \ {\omega}<{\omega}_{0} \ {\mbox{or}} \ 
     {\omega}>{\omega}_{0}+{\Delta\omega},
\end{array} }
\end{equation}
with the step width $\Delta\omega$, which is normally much 
greater than the width of the high--frequency vibration band. 
In addition, suppose resonance widths ${\eta}_{\bf K}(\omega)$ 
to be much the same on the interval $\Delta\omega \ 
\left [ {\eta}_{\bf K}(\omega) \approx 
{\eta}_{\bf K}({\omega}_{0}) \equiv \eta \right ]$. 
Then, relation (27) takes the form:

\begin{equation}
W=\frac{\gamma}{\Delta\Omega}e^{-\hbar\omega_{0}/k_{B}T}
\int\limits_{0}^{\Delta\omega}dxe^{-\hbar x/k_{B}T}
\left ( 1+\frac{\gamma}{\Delta\Omega}
\ln\frac{x+i\eta-\Delta\omega}{x+i\eta} \right )^{-1}.
\end{equation}
Assuming the sufficiently small ratio $\Delta\omega/\eta$ and 
allowing for the fact that the quantities $\Delta\Omega$ and 
$2\Gamma$ essentially differ from zero only when $k_{B}T$ 
approaches $\omega_{0}$ (i.e., provided the inequalities 
$\Delta\omega << \eta << k_{B}T /\hbar < \omega_{0}$  hold), 
we obtain, accurate to terms of the orders 
$\Delta\omega/T$ and $(\Delta\omega/\eta)^{2}$:

\begin{eqnarray}
\Delta\Omega & \approx & \frac{\gamma\eta^{2}}{\gamma^{2}+\eta^{2}}
\left [ 1-\frac{\hbar\Delta\omega}{2k_{B}T}+
\frac{\gamma^{2}\left (9\gamma^{2}+13\eta^{2}\right ) }
     {12\left (\gamma^{2}+\eta^{2}\right )^2}
\left ( \frac{\Delta\omega}{\eta} \right )^{2} \right ]
e^{-\hbar\omega_{0}/k_{B}T},\\
2\Gamma & \approx & \frac{2\gamma^{2}\eta}{\gamma^{2}+\eta^{2}}
\left [ 1-\frac{\hbar\Delta\omega}{2k_{B}T}-
\frac{6\eta^{4}-3\gamma^{2}\eta^{2}-5\gamma^{4}}
     {12\left (\gamma^{2}+\eta^{2}\right )^2}
\left ( \frac{\Delta\omega}{\eta} \right )^{2} \right ]
e^{-\hbar\omega_{0}/k_{B}T}.
\end{eqnarray}

At $\Delta\omega=0$, relations (32) and (33) are reduced 
to the reported results [8] for an adsorbed molecule. 
Lateral interactions of adsorbed molecules give rise to 
the low--frequency molecular band with the non--zero 
parameter $\Delta\omega$. The integration over this band 
leads to the decreased shift and width of the high--frequency 
spectral line. Since the parameters $\hbar\Delta\omega/k_{B}T$ 
and $(\Delta\omega/\eta)^{2}$ in the temperature range of 
interest can be of the same order of magnitude, the contribution 
$(\Delta\omega/\eta)^{2}$ to the value $\Delta\Omega$ can be 
only positive, whereas the sign of this contribution to 
the value $2\Gamma$ depends on the $\eta$ to $\gamma$ ratio. 
At $\eta/\gamma > 1.094$ (which is consistent with the realistic 
values of $\eta$ and $\gamma$), the contribution 
$(\Delta\omega/\eta)^{2}$ causes the additional narrowing 
of the spectral line.

Table 1 shows temperature dependences of the high--frequency 
line shift and width, with the parameter values corresponding 
to isotopically diluted {$^{13}$}C{$^{16}$}O{$_{2}$} molecules 
in the {$^{12}$}C{$^{16}$}O{$_{2}$} monolayer on the NaCl(100) 
surface~[25]. The data outlined are indicative of changes 
in spectral line shifts and widths when passing from the band 
width $\Delta\omega = 5$~cm{$^{-1}$} to $\Delta\omega = 0$. 
These dependences evidently demonstrate the narrowing of 
the spectral line, in accordance with experimental results~[25]. 
In the framework of the conventional exchange dephasing model, 
i.e., with no regard to lateral interactions of low--frequency 
modes, an attempted description of the temperature dependences 
for the above-mentioned {$^{13}$}C{$^{16}$}O{$_{2}$} molecular 
ensemble resulted in overestimated line widths, whereas 
theoretically calculated line shifts agreed well with experimental 
values. Stretch vibration frequencies for 
{$^{13}$}C{$^{16}$}O{$_{2}$} and {$^{12}$}C{$^{16}$}O{$_{2}$} 
molecules differ by 60~cm{$^{-1}$} and their coupling can thus 
be neglected. In contrast, low--frequency vibrations prove 
to be essentially coupled: As far as translational vibrations 
are concerned, the mass difference between carbon isotopes is 
slight compared to the total molecular mass, whereas for 
orientational vibrations, the mass of the central carbon atom 
has only a slight effect on the molecular moment of inertia. 
With the coupling of low--frequency vibrations included, 
the model suggested provides an accordance between 
calculated and observed spectral line widths. 

        This research was supported by the State Found of 
Fundamental Researches administered by the State Committee 
on Science and Technology of Ukraine (Project 2.4/308).

\newpage

Table 1.  Temperature dependences of the spectral line shift (a) 
and width (b) for molecular local vibrations, with intermolecular 
interactions neglected ($\Delta\omega=0$) and included 
($\Delta\omega=5$~cm{$^{-1}$}) for the following parameter values: 
$\gamma=0.7$~cm{$^{-1}$}, $\eta=1.2$~cm{$^{-1}$}, 
${\omega}_{0}=41$~cm{$^{-1}$}.

\begin{center}
{\large (a)}\\
\vspace{5mm}
\begin{tabular}{|c|p{5cm}|p{5cm}|}
\hline
T (K) & \hspace{15mm}$\Delta\Omega$ (cm{$^{-1}$}) 
        & \hspace{15mm}$\Delta\Omega$ (cm{$^{-1}$}) \\ 
      & \hspace{15mm}at $\Delta\omega=0$          
        & \hspace{11mm}at $\Delta\omega=5$~cm{$^{-1}$}\\ 
\hline
20 & \hspace{17mm}0.027223 & \hspace{17mm}0.029096 \\
\hline
30 & \hspace{17mm}0.072878 & \hspace{17mm}0.082131 \\
\hline
40 & \hspace{17mm}0.119241 & \hspace{17mm}0.138051 \\
\hline
50 & \hspace{17mm}0.160221 & \hspace{17mm}0.188547 \\
\hline
60 & \hspace{17mm}0.195096 & \hspace{17mm}0.232113 \\
\hline
70 & \hspace{17mm}0.224566 & \hspace{17mm}0.269276 \\
\hline
80 & \hspace{17mm}0.249553 & \hspace{17mm}0.301008 \\
\hline
\end{tabular}

\end{center}

\vspace{10mm}

\begin{center}

{\large (b)}\\
\vspace{5mm}
\begin{tabular}{|c|p{5cm}|p{5cm}|}
\hline
T (K) & \hspace{17mm}$2\Gamma$ (cm{$^{-1}$}) 
        & \hspace{17mm}$2\Gamma$ (cm{$^{-1}$}) \\ 
      & \hspace{15mm}at $\Delta\omega=0$          
        & \hspace{11mm}at $\Delta\omega=5$~cm{$^{-1}$}\\ 
\hline
20 & \hspace{17mm}0.031761 & \hspace{17mm}0.016631 \\
\hline
30 & \hspace{17mm}0.085025 & \hspace{17mm}0.046801 \\
\hline
40 & \hspace{17mm}0.139113 & \hspace{17mm}0.078537 \\
\hline
50 & \hspace{17mm}0.186925 & \hspace{17mm}0.107158 \\
\hline
60 & \hspace{17mm}0.227612 & \hspace{17mm}0.131831 \\
\hline
70 & \hspace{17mm}0.261993 & \hspace{17mm}0.152864 \\
\hline
80 & \hspace{17mm}0.291145 & \hspace{17mm}0.170817 \\
\hline
\end{tabular}

\end{center}

\end{document}